\documentclass[final]{jfp-epi}

\usepackage{graphicx}
\usepackage{svg}

\usepackage{listings}
\usepackage{xcolor}
\definecolor{commentgreen}{RGB}{2,112,10}
\definecolor{eminence}{RGB}{108,48,130}
\definecolor{weborange}{RGB}{255,165,0}
\definecolor{frenchplum}{RGB}{129,20,83}
\jfpVolume{36}
\jfpArticle{11}
\jfpDOI{10.46298/jfp.17822}
\jfpYear{2026}
\received[Submitted]{February 2025}
\received[accepted]{March 2026}

\begin{document}

\title{Adapting the MVVM pattern to C++ frontends and Agda-based backends}

\author{Viktor Csimma}
\orcid{1234-5678-9012}
\affiliation{%
  \institution{Eötvös Loránd University, Faculty of Informatics and Eötvös József Collegium}
  \city{Budapest}
  \country{Hungary}
  \authoremail{csimmaviktor03@gmail.com}
}

\begin{abstract}

Using agda2hs and ad-hoc Haskell FFI bindings,
writing Qt applications in C++
with Agda- or Haskell-based backends (possibly including correctness proofs)
is already possible.
However, there was no repeatable methodology to do so,
nor to use \emph{arbitrary} Haskell built-in libraries
in Agda code.

We present a well-documented, general methodology
to address this,
applying the ideas of the Model-View-ViewModel architecture
to models implemented in functional languages.
This is augmented by
a software development kit
providing easy installation and automated compilation.

For obstacles arising,
we provide solutions and ideas
that are novel contributions by themselves.
We describe and compare solutions
for using arbitrary Haskell built-ins
in Agda code,
highlighting their advantages and disadvantages.

Also, for user interruption,
we present a new Haskell future design.
This, to the best of our knowledge,
is the first to provide for arbitrary interruption
and the first to provide for interruption
via direct FFI calls from C and C++.

Finally, we prove with benchmarks that
the agda2hs compiler at the base of our methodology
is viable when compared to other solutions,
specifically to the OCaml extraction feature of Rocq
and the default MAlonzo backend of Agda.

\end{abstract}

\maketitle

\section{Introduction}

For decades, verifying computer programs,
even in functional programming,
has been comparable to
thoroughly reading all the contracts
we sign at a bank:
everyone agrees it is essential,
but no one really has the time
to actually do so.
Unable to solve this problem,
we wish to at least lower the threshold
for developers or companies thinking about
writing a functional backend
and verifying parts of it.

First, we enumerate our contributions,
as well as the key challenges they solve.
We also provide some technical information
that may be necessary for understanding the article.

\subsection{Contributions}
\label{subsec:contributions}

More than 15 years after the introduction of the current iteration of Agda,
the \emph{agda2hs} compiler \citep{agda2hs-paper} came out,
providing a way to generate human-readable Haskell code of reasonable speed
from specially formatted Agda programs.
In previous papers,
we have already demonstrated the practical usability of agda2hs
by building an exact-real arithmetic library called \emph{Acorn} \citep{acorn-article}
and then connecting it to a Qt-based C++ frontend named \emph{AcornCalc} \citep{calc-article}.
Still, the solutions presented there for using Haskell libraries
or connecting a C++ frontend
were very problem-specific,
rather than parts of a repeatable methodology.

In this paper, a generic approach is presented
to tackle the above two challenges,
along with an entire development kit
(named \emph{Agdalache})
which simplifies the process,
enabling Agda developers to easily add GUI frontends
to their applications.

Our contributions are the following,
with the motivation behind them explained
in Section \ref{subsec:motivation}:

\begin{itemize}
\item A general methodology and workflow
  to build backends in Agda
  and C/C++ GUI frontends running on top of them,
  as demonstrated in Section \ref{sec:methodology}
  and \ref{subsec:appstate}.
  While this is what the framework itself is based on,
  we would like to stress that this is not just a software-engineering project:
  rather, it should be interpreted as a set of design patterns and best practices,
  notably including the adaptation of
  the Model-View-ViewModel architecture
  to backends written in functional languages
  (which may be relevant outside the scope of Haskell and Agda as well).
\item A precise description and comparison
  of different methods by which
  \emph{arbitrary} GHC Haskell built-in libraries
  can be used in Agda code,
  together with their thorough comparison
  and a reasoning why postulates are currently the best choice
  for a software development kit like ours.
  These are described in Section \ref{subsec:haskell-dependencies}.
\item A new design of \emph{futures} in Haskell
  (and, by extension, in Agda)
  that is interruptible anytime,
  both from Haskell and from a C/C++ frontend
  (for the latter, even a RAII class is presented).
  To the best of our knowledge,
  this is the first future concept in Haskell
  capable of arbitrary interruption,
  let alone of being interrupted from C++
  via a native FFI function
  instead of a full Haskell call.
  For details, see Section \ref{subsec:futures}.
\item A software development kit with pre-written primitives for MVVM,
  as well as build scripts
  that can compile Agda/Haskell code
  together with complex C++ Qt frontends.
  While this may be perceived as a software engineering feat
  rather than a scientific contribution,
  we do not know of any previous attempts
  to tackle the problem of automated compilation
  in such projects.
\item A demonstration, by benchmarks,
  that the performance of agda2hs-based programs
  is viable when compared to Rocq-based solutions.
  This establishes the viability of all the other results of this paper as well.
  See Section \ref{sec:benchmarks}.
\end{itemize}

\subsection{Motivation}
\label{subsec:motivation}

Here, we highlight the main problems we have to overcome
when writing an Agda-based backend with a C/C++ frontend,
as well as how the corresponding contributions solve them.

\subsubsection{Modelling Haskell libraries}

As Agda code is eventually compiled by agda2hs to plain Haskell,
one probably wants to use the built-in libraries provided by that language,
instead of, say, the Agda standard library
(which is less efficient and
mostly not agda2hs-compatible anyway).
The original solution for that is
the standard library distributed with agda2hs
\citep[folder \texttt{lib}]{agda2hs},
which can be imported and referred to in Agda code,
but whose concepts are recognised by the compiler and
automatically translated to the original Haskell counterparts.

That library, however, is not comprehensive:
for a long time, only base concepts of the Prelude
and monad-related libraries had been implemented;
although some containers (Data.Map, Data.Set) were added in the summer of 2025
\citep[commit \texttt{cad343f}]{agda2hs}.
Therefore, one may wonder what to do
if something outside the library concept is needed,
like Data.IORef---not to mention FFI-related modules,
e.g.\ conversion functions from the Foreign.C module.

In Section \ref{subsec:haskell-dependencies},
we are going to present different solutions to this problem,
present their advantages and disadvantages,
and provide examples for their usage in related projects.
We do not know of any previous systematic examination
of such approaches.

\subsubsection{Background calculations}

During our work, we have to solve the problem of
making background calculations easily interruptible,
as in a GUI application,
the user typically expects a mechanism
through which they can interrupt a process
that seems to be too long.
For our target group (i.e.\ Agda developers connecting GUIs to their programs),
it is especially important to make this simple,
so that they can concentrate on proving the correctness of business logic
and need not struggle with long and hardly verifiable C++ constructions.

Although futures have already existed in C++
and even in Haskell \citep{old-future},
no Haskell futures have been created
that can be interrupted unconditionally with a single function call,
let alone from the other side of the FFI.

Our contribution is
a new design of futures
described in Section \ref{subsec:futures},
based on two MVars
and a ``watcher thread'' that interrupts the calculation thread when activated.
The implementation puts a special emphasis on interruptibility
and connection with C/C++ frontends,
with Future objects easy to export and manipulate from the outside---%
especially as interruption can be done via a native function of the FFI,
without a full Haskell call.

As it is not Agda-specific
and might also be useful for ``ordinary'' Haskell projects,
the implementation has been made available separately on Hackage
by the name \href{https://hackage.haskell.org/package/cfuture}{cfuture}.
In Agdalache, we basically provide an Agda interface to this library.

\subsubsection{Ease of compilation}

GHC allows the developer to compile C and C++ code
and link it to the Haskell parts of the application
\citep{ghc-docs-ffi}.
However, when coming to an entire frontend project
where listing all C/C++ source files is unfeasible,
it becomes hard to stick with GHC:
build systems (specifically, CMake) assume a ``usual'' C compiler
like GCC or Clang,
but Clang does not include all the necessary libraries by itself,
so they have to be specified in a build script.

We provide such pre-written build scripts
so that developers do not need to hassle
with these technicalities.

\subsubsection{Ease of installation}

The standard package management system of Haskell
(and, by extension, Agda)
is Cabal \citep{cabal-docs-what-cabal-does}.
It is, however, hard to use in Agda-related projects.
The author can justify this harsh claim
by referring to his 2-year experience with teaching Agda;
as in each semester, the majority of students
could not even install the type checker itself
without technical problems and external troubleshooting,
in spite of clear instructions---%
the most common problems were dependency resolution errors
and missing dependencies from outside the Haskell ecosystem
(usually zlib).
The situation becomes even worse when turning to less established packages,
like agda2hs.

\subsubsection{Demonstration of viability}

Given that the Rocq proof assistant already provides solutions
for code extraction,
the mere existence of the project is justified
only if we can demonstrate that
using agda2hs is also a viable way of
writing programs of a practically usable speed.
We do so in Section \ref{sec:benchmarks},
via benchmarks which show that agda2hs-produced programs
have a speed comparable to Rocq code
translated to OCaml.
From here on, as Agda developers probably prefer
the syntax and philosophy of Agda to those of Rocq, 
they can choose agda2hs (and thus Agdalache) without worries.

\subsection{Necessary pre-knowledge}

Along with a general knowledge of Agda, Haskell and C++,
the following concepts are needed
to understand the technical details of this paper:
\begin{itemize}
\item In Haskell, an \textbf{MVar} (\texttt{MVar t}) is a mutable location
  that is either empty
  or contains a value of type t.
  It also functions as a semaphore:
  \texttt{putMVar} fills an MVar if it is empty and waits otherwise until it gets empty;
  \texttt{takeMVar} empties an MVar if it is filled and waits otherwise until it gets filled
  \citep[Control.Concurrent.MVar]{base}.
  Of course, all such operations live in the IO monad.
\item A \textbf{StablePtr} (\texttt{StablePtr t}) is an identifier referring to a Haskell object;
  it can be exported to C and passed back to a backend call
  when a specific object has to be reached.
  Until the StablePtr is freed, the object is guaranteed
  not to be deallocated by the garbage collector.
  Technically, a StablePtr is represented as a \texttt{void*} in C,
  but it is usually not an actual memory address;
  rather an index (e.g. 4)
  for a lookup table managed by the Haskell runtime
  \citep[Foreign.StablePtr]{base}.
\item \textbf{``Resource acquisition is initialisation'' (RAII)} is a term
  coined by Bjarne \citet{cpp-evolution}
  and usually used for a programming practice in the C++ language.
  The basic idea is that if we have to initialise and later release a resource,
  we represent it with a local object,
  so that its destructor also releases the underlying resource.
  This way, for example,
  resources held by an object on the execution stack
  immediately get freed,
  no matter how the containing function exits
  \citep{stroustrup-faq}.
  It is also useful for objects allocated on the heap,
  as we do not need to individually release each resource
  it has acquired.
\item The \textbf{``Model-View-ViewModel'' (MVVM)} architecture
  is a design pattern for GUI applications,
  designed by Ken Cooper and Ted Peters at Microsoft
  for the Windows Presentation Foundation framework
  and first publicly described in 2005 by John \citet{mvvm-2005}.
  In this pattern, the developer splits responsibilities
  between three layers:
  the model contains business logic,
  the view model is responsible for presentation logic
  (e.g.\ values that are to be given to text boxes)
  \emph{without} actually depending on GUI libraries,
  and the view does the actual presentation
  via the GUI capabilities of the framework
  \citep{mvvm-book}.
  This way, the portability and flexibility of the program greatly increases,
  as all three layers can be swapped independently of the other
  if provided the same API,
  and for a different framework, only the view needs to be updated.

  MVVM is widely used outside the .NET ecosystem as well,
  and the primitives of Agdalache are built to support this pattern,
  with the model being implemented on the Agda/Haskell side
  and exported.
  (However, the model object can actually be used
  no matter what design choices the developer makes
  when writing the frontend.)
\end{itemize}

\section{Background -- Agda and agda2hs}
\label{sec:background}

The original Agda project was started in 1999
\citep{agda1-paper}.
An interactive interface was soon presented for it,
followed by an Emacs extension \citep{agda-emacs-paper}.
In 2005, when Ulf Norell and Andreas Abel started
a complete rewrite called Agda 2,
their goal was to create
``a practical programming language based on dependent type theory'' \citep{norell-thesis}.
Today, Agda 2 is simply referred to as ``Agda'',
having succeeded the previous implementation.

In accordance with the goal of practical usefulness,
a compiler called \emph{MAlonzo},
also referred to simply as ``the GHC backend'' \citep[Section ``Compilers'']{agda-docs},
has been bundled with the Agda type checker
since at least 2008 \citep{malonzo-commit}.
This translates any legal Agda code
to Haskell,
with type coercions being used
to simulate dependent types
(since they are not available in Haskell).
As this creates obfuscated, hard-to-read code,
it is usually not enough
for integrating Agda code
into a Haskell environment:

\vspace{5mm}

\emph{``While many of the coercions
inserted by MAlonzo are not necessary to make the
code be accepted by GHC, they cannot be avoided in general
because Agda supports full dependent types, while GHC
(currently) does not. The coercions inserted by MAlonzo
make it difficult to make the jump from having an Agda
prototype of a Haskell program (or a component in a larger
Haskell project) to having a production Haskell program.''}---\citet{agda2hs-paper}

\vspace{5mm}

This was one of the reasons
why \emph{agda2hs}
was introduced in 2022 \citep{agda2hs-paper}.
agda2hs is a code extraction tool
(although throughout this paper,
we are going to refer to it as a compiler too),
similar to those built into
Rocq \citep[Section ``Program extraction'']{rocq-reference-manual}
and other theorem provers.
Since dependent types are not supported in Haskell,
agda2hs operates only on a subset of Agda
where all dependently-typed portions
have been marked for the compiler as irrelevant—%
such marks are referred to as \emph{erasure annotations}
\citep[Section ``Run-time Irrelevance'']{agda-docs}.
However, it generates human-readable Haskell code
that can be understood even by developers with no knowledge of Agda,
thereby enabling users to include verified Agda code
in a Haskell project.

Translation to human-readable Haskell code
should also logically imply
a speed equivalent with that of handwritten Haskell.
Still, to correctly evaluate the practical usefulness
of the development kit presented here,
we have to examine
whether agda2hs is fast enough
to be viable among other solutions---%
particularly in comparison to
the code extraction capabilities
of Rocq.

\emph{Rocq} (formerly called \emph{Coq} until March 2025)
is also an interactive theorem prover,
one of the most well-known systems of its kind,
with notable applications
including a verified proof of the four-colour theorem \citep{gonthier}
and the CompCert C compiler \citep{compcert-paper}---%
the latter even having industrial applications,
with an extended version being sold as a commercial product.

Rocq has a built-in tool for extracting definitions
to lower-level functional languages,
with the default being OCaml
(Haskell and Scheme are also supported).
Also, recently, a verified foreign interface between Rocq and C,
called VeriFFI, has been announced by \citet{veriffi},
with a correctness proof in VST and Rocq itself.
Due to its establishment and continuously developing ecosystem,
we are going to use Rocq as reference
when assessing the viability of agda2hs-based solutions
(e.g.\ in benchmarks).

\section{Repositories and self-references}
\label{sec:repos-and-links}

This article focuses on
the scientifically relevant aspects of our work;
namely, the repeatable methodology,
the novel patterns and designs introduced,
and the proof of viability by benchmarks.
For technical details
(e.g.\ instructions and implementation techniques),
we refer the reader to
\href{https://agdalache.readthedocs.io/en/latest/}{the documentation}
of Agdalache,
as well as the code itself.

We also provide a list of
all the self-made GitHub repositories referred to later on,
as well as previous articles:

\begin{itemize}
\item The repository of the SDK,
  containing the skeleton classes described
  as well as the build and installation scripts,
  is \href{https://github.com/viktorcsimma/agdalache}{viktorcsimma/agdalache}.
\item The EvenCounter example project,
  described in Section \ref{subsec:evencounter},
  can be found at \href{https://github.com/viktorcsimma/even-counter}{viktorcsimma/even-counter}.
  A Windows binary can also be downloaded there.
\item The cfuture Haskell package,
  which makes the exportable futures available for any Haskell project,
  has been published on \href{https://hackage.haskell.org/package/cfuture}{Hackage}.
  Agdalache provides an Agda wrapper for the 2.0 version of this library.
\item The benchmark files used in Section \ref{sec:benchmarks}
  are under \href{https://github.com/viktorcsimma/benchmarks/}{viktorcsimma/benchmarks}.
\item The AcornCalc exact-real calculator,
  presented in Section \ref{subsec:acorn},
  is divided into two separate repositories:
  \begin{itemize}
  \item the backend, called Acorn, is at
    \href{https://github.com/viktorcsimma/acorn}{viktorcsimma/acorn};
  \item while the frontend is at
    \href{https://github.com/viktorcsimma/acorn-calc}{viktorcsimma/acorn-calc}
    (here, Linux and Windows binaries are also included).
  \end{itemize}
  The two corresponding articles are \citet{acorn-article} and \citet{calc-article}.
\end{itemize}

\section{The MVVM-based methodology proposed}
\label{sec:methodology}

Here, we present the methodology in detail,
walking through how an example project is built up
from verified business logic to the C++ level,
and also having a look at a larger project
relying on the same principles.

\subsection{EvenCounter}
\label{subsec:evencounter}

\begin{figure}[b]
\begin{center}
\includegraphics[scale=.4]{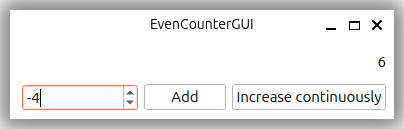}
\includegraphics[scale=.4]{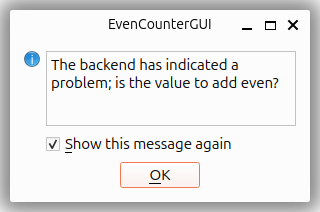}
\caption{Screenshots of EvenCounter.}
\label{fig:even-counter}
\end{center}
\end{figure}

\emph{EvenCounter} is a small demo program,
used here for demonstrating
a practical usage of the concepts introduced.
The idea is having a counter
that is proven in Agda to be able to contain
only even numbers.
Concrete values can be added to the current one
from the GUI,
and another button increases the counter with 2
every second, for five seconds,
in an interruptible way.

See Figure \ref{fig:even-counter} for screenshots.

\subsubsection{Verified code}

In \texttt{Logic.agda},
we prove that the sum of two even integers
is also even.
(The details of this are not needed to follow along,
but it starts with pattern matching on
whether the parameters are nonnegative or negative.)

\begin{lstlisting}[language=Haskell]
@0 intSumIsEven : (a b : Integer)
                  -> @0 {EvenInteger a} -> @0 {EvenInteger b}
                  -> EvenInteger (a + b)
intSumIsEven (pos m) (pos n) {em} {en} = ...
intSumIsEven (pos zero) (negsuc (suc n)) {_} {en} = ...
...
\end{lstlisting}

We then write a function which only takes two even integers,
and returns the sum along with the proof that it is even,
in a sigma type \texttt{S0}.
In this type, the second part of the tuple (usually a proof) is erased,
and only the first object remains at runtime.
(For technical reasons, the operator \texttt{:\&:}
is represented here as \texttt{:\textasciicircum:}.)

\begin{lstlisting}[language=Haskell]
addEvenIntegers : (x y : Integer)
    -> @0 {EvenInteger x} -> @0 {EvenInteger y}
    -> S0 Integer EvenInteger
addEvenIntegers x y {ex} {ey}
  = (x + y) :^: intSumIsEven x y {ex} {ey}
{-# COMPILE AGDA2HS addEvenIntegers #-}
\end{lstlisting}

Finally, we call this from a function that does not take proofs,
so that it can be used in Haskell.
It returns an error message in an Either instead
if either of the parameters are even.
Here, we would like to stress that
the function passes the type checker only because
\texttt{addEvenIntegers} got proofs for the evenness of
both \texttt{x} and \texttt{y};
were we to make a mistake
(e.g.\ by using a wrong branch of the if structures),
type checking would fail.

\begin{lstlisting}[language=Haskell]
eitherAddInteger : Integer -> Integer -> Either String Integer
eitherAddInteger x y =
  if (isEvenInteger x) then (\ {{isTrue1}} ->
    if (isEvenInteger y) then (\ {{isTrue2}} ->
      Right (proj1 (
              addEvenIntegers x y {equivToEvenInteger {x} isTrue1}
                                  {equivToEvenInteger {y} isTrue2}
      ))
    ) else Left "second parameter is odd"
  ) else Left "first parameter is odd"
{-# COMPILE AGDA2HS eitherAddInteger #-}
\end{lstlisting}

As the first function is only a proof and is marked with \texttt{@0},
only the second two are extracted to Haskell.
As the second parts of \texttt{S0} sigmas are erased,
the result becomes simple:

\begin{lstlisting}[language=Haskell]
addEvenIntegers :: Integer -> Integer -> S0 Integer
addEvenIntegers x y = ((x + y) :^:)

eitherAddInteger :: Integer -> Integer -> Either String Integer
eitherAddInteger x y
  = if isEvenInteger x then
      if isEvenInteger y
      then Right (proj1 (addEvenIntegers x y))
      else Left "second parameter is odd"
    else Left "first parameter is odd"
\end{lstlisting}

\subsubsection{Backend}
\label{subsubsec:evencounter-backend}

To create a GUI application from the above functions,
we first define a type called AppState.
In our case, it contains an IORef to an integer;
this is going to be the actual value of the counter
which can always be modified via the same AppState instance.
For a longer explanation on why it is defined this way,
see Section \ref{subsec:appstate}.

\begin{lstlisting}[language=Haskell]
record AppState : Set where
  constructor MkAppState
  field
    counterRef : IORef Integer
open AppState public
{-# COMPILE AGDA2HS AppState #-}
\end{lstlisting}

Initialisation of a concrete object goes like this:

\begin{lstlisting}[language=Haskell]
-- An AppState initialised with a given number.
initAppState : Integer -> IO AppState
initAppState n = MkAppState <$> newIORef n
{-# COMPILE AGDA2HS initAppState #-}
\end{lstlisting}

In \texttt{Interaction.agda}, we then proceed to create a function
that is exportable to C
(always setting 0 as the initial value,
for the sake of simplicity):

\begin{lstlisting}[language=Haskell]
initApp : IO (StablePtr AppState)
initApp = newStablePtr =<< initAppState 0
{-# COMPILE AGDA2HS initApp #-}

{-# FOREIGN AGDA2HS
foreign export ccall initApp :: IO (StablePtr AppState)
#-}
\end{lstlisting}

Incrementing the counter, using the verified functions,
is done by a separate function taking the AppState object:

\begin{lstlisting}[language=Haskell]
incrementInteger' :
    AppState -> Integer -> IO (Either String Integer)
incrementInteger' appState x = do
  inner <- readIORef (counterRef appState)
  let either = eitherAddInteger inner x
  case either of \ where
    (Left err) -> return either -- pass the error message
    (Right result) -> do
      writeIORef (counterRef appState) result
      return either
{-# COMPILE AGDA2HS incrementInteger' #-}
\end{lstlisting}

And after some type conversions, this can be exported to C
as a synchronous call.

\subsubsection{Asynchronous calculation}

Now, let us suppose we have a function
that increments the value every second by 2,
stopping after a given duration:

\begin{lstlisting}[language=Haskell]
{-# TERMINATING #-}
increaseContinuouslyInteger : AppState Integer -> Int -> IO Integer
increaseContinuouslyInteger appState duration =
  if 0 < duration then (do
              threadDelay 1000000
              incrementInteger' appState 2
              print =<< readIORef (counterRef appState)
              increaseContinuouslyInteger appState (duration - 1))
  else readIORef (counterRef appState)
{-# COMPILE AGDA2HS increaseContinuouslyInteger #-}
\end{lstlisting}
To create an exportable definition from this,
we call some auxiliary functions:
\begin{lstlisting}[language=Haskell]
increaseContinuouslyIntegerAsyncC
   : StablePtr (AppState Integer) -> Int -> CFuturePtr -> IO T
increaseContinuouslyIntegerAsyncC appStatePtr duration futurePtr
  = forkFutureC futurePtr
    $ (deRefStablePtr appStatePtr) >>= (\ appState ->
        unsafeIntegerToCInt <$>
                  increaseContinuouslyInteger appState duration)
{-# COMPILE AGDA2HS increaseContinuouslyIntegerAsyncC #-}

{-# FOREIGN AGDA2HS
foreign export ccall increaseContinuouslyIntegerAsyncC ::
  StablePtr (AppState Integer) -> Int -> CFuturePtr -> IO ()
#-}
\end{lstlisting}

Most of this code is only about conversions;
the important part is calling \texttt{forkFutureC}
on the calculation \texttt{increaseContinuouslyInteger appState duration},
thus writing a future to the C pointer \texttt{futurePtr}.

This function can now be exported to C in a header file, as:

\begin{lstlisting}[language=C]
extern void increaseContinuouslyIntegerAsyncC(
    HsStablePtr appState, int duration, HsStablePtr* future
);
\end{lstlisting}

\subsubsection{C++}

On the C++ side, we provide a class called HsAppStateWrapper;
this is going to hold a StablePtr to the AppState object above,
in a RAII way.
It basically represents the model towards the upper layers.
Again, for more elaboration on this,
see Section \ref{subsec:appstate}.

\begin{lstlisting}[language=C]
class HsAppStateWrapper {
  private:
    const HsStablePtr appStatePtr;

  public:
    HsAppStateWrapper():
      // initApp :: IO ( StablePtr AppState )
      appStateWrapper(initApp()) {}
  
    bool incrementWith(int toAdd) {
      // incrementInteger ::
      //    StablePtr (AppState Integer) -> CInt -> IO CInt
      // see the code for the definition
      return incrementInteger(appStatePtr, toAdd);
    }
    
    ~HsAppStateWrapper() {
      // a built-in FFI call
      hs_free_stable_ptr(appStatePtr);
    }

    // this is going to call the previous Agda definition
    Future<int> increaseContinuouslyAsync(int duration);
};
\end{lstlisting}
The last method is defined as: 
\begin{lstlisting}[language=C]
Future<int>
    HsAppStateWrapper::increaseContinuouslyAsync(int duration) {
  return Future<int>([this, duration](HsPtr futurePtr){
    // increaseContinuouslyIntegerAsyncC ::
    //    StablePtr (AppState Integer) -> Int -> CFuturePtr -> IO ()
    increaseContinuouslyIntegerAsyncC(
        appStatePtr, duration, futurePtr
    );
  });
}
\end{lstlisting}
And from now on, the calculation can be started any time:
\begin{lstlisting}[language=C]
Future<int> future = appStateWrapper.increaseContinuouslyAsync(42);

// And if we want to interrupt:
future.interrupt();
\end{lstlisting}
Finally, we use the objects
just as we would in any C++ project,
using them as the way of communicating with the model.
In our projects, the GUI is based on Qt.

\subsection{Acorn and AcornCalc}
\label{subsec:acorn}

\begin{figure}[h]
\begin{center}
\includegraphics[height=45mm]{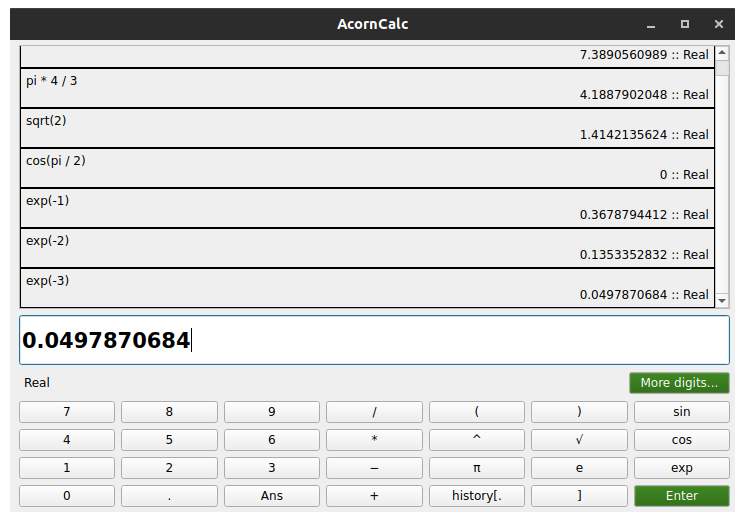}
\includegraphics[height=45mm]{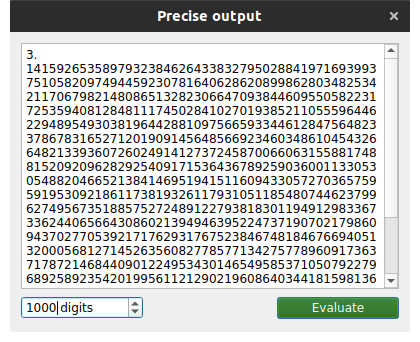}
\caption{Screenshots of AcornCalc.}
\label{fig:acorn-calc}
\end{center}
\end{figure}

Acorn is a previous project
which is an example of a practical, large-scale use case:
it embeds exact-real arithmetic into
a GUI calculator application.
See Figure \ref{fig:acorn-calc} for screenshots.
Notably, it does \emph{not} use the Agdalache framework
(it is rather a precursor to it),
but the methodology is very similar.

The project actually consists of two repositories;
the Acorn backend (written in Agda and Haskell)
and the AcornCalc frontend (written in C++).
For links and previous articles, we refer to Section \ref{sec:repos-and-links}.

As the focus of the project was mostly on the arithmetic itself,
and the C++ connection is rather a predecessor of the final SDK,
we are only going to take a look at some parts.
The AppState type (here called \emph{CalcState})
is defined in a way roughly equivalent
to the following code in Haskell:

\begin{lstlisting}[language=Haskell]
data CalcState real = MkCalcState
  { variables :: IORef (Map String (Value real))
  , history   :: IORef [Value real]
  }
\end{lstlisting}

Here, \texttt{real} is a type variable
(that can be substituted for two different implementations);
this demonstrates how type variables can be used
in the AppState definition.

The value manipulation functions consist mostly of parsing and interpreting
expressions given by the user,
which is outside the scope of this paper.
However, Acorn also includes a command prompt
written in Haskell,
whose main function is:

\begin{lstlisting}[language=Haskell]
main :: IO ()
main = do
  putStrLn "Welcome to the AcornShell interpreter."
  calcState <- emptyCalcState
  -- 'prompt' is a recursive function here,
  -- corresponding to a single value entry
  prompt calcState 100

emptyCalcState :: IO (CalcState real)
emptyCalcState = do
  varsRef <- newIORef Map.empty
  histRef <- newIORef []
  return $ MkCalcState varsRef histRef
\end{lstlisting}

In short, after initialising a CalcState object
(only once, at the very beginning of the program)
with empty variables,
we pass this to the \texttt{prompt} function
which retains and handles it throughout recursive calls.

Now, back to the GUI connection.
In Shell.Interaction, a function is defined
that takes a command in a C string format,
executes it
and converts the result back to a C string.
(The precision parameter \texttt{prec}
is an Acorn-specific concept;
it is the number of precise decimal digits
we want to have in the result.)

\begin{lstlisting}[language=Haskell]
execCommand :: StablePtr (CalcState real) -> CString -> CInt
               -> IO CString
execCommand ptr cstr prec = do
  command <- peekCString cstr
  calcState <- deRefStablePtr ptr
  answer <- execCommand' calcState command (fromIntegral prec)
  newCString answer
\end{lstlisting}

This can then be exported to C---%
however, as the type variable \texttt{real} has to be eliminated,
we actually export two functions:
\texttt{execCommandDyadic} for \texttt{CalcState Dyadic}
and \texttt{execCommandRational} for \texttt{CalcState Rational}.
The resulting type signature for the former is:

\begin{lstlisting}[language=C]
extern char* execCommandDyadic(
  HsStablePtr calcState, char* command, int precision
);
\end{lstlisting}

And in HsCalcStateWrapper, the functions are called like this:

\begin{lstlisting}[language=C]
std::string HsCalcStateWrapper::execCommand
  (const char* command, int precision) const
{
  // we check which implementation to use
  const char* result
    = (DyadicBase == baseType)
    ? (char*) execCommandDyadic(
          calcStatePtr, (char*) command, precision
      )
    : (char*) execCommandRational(
          calcStatePtr, (char*) command, precision
      );
  // we copy it into a string
  std::string toReturn(result);
  // the C string can be freed from here,
  // even though it has been created by the backend
  free((void*) result);
  // and we hope for a return-value optimisation
  return toReturn;
}
\end{lstlisting}

For background calculations, AcornCalc uses OS semaphores instead of futures;
futures are one of the new features the current SDK has brought about.

For more details, see the papers directly related to the program
\citep{acorn-article,calc-article}.

\section{Implementation-related contributions}

Here, we go into more details about
the implementation of the framework,
which contains contributions
that are notable by themselves.

\subsection{AppState}
\label{subsec:appstate}

On the Agda/Haskell side,
the template provides a type called AppState ---%
this is basically to be the \emph{model} of our MVVM application,
implemented entirely in the backend layer.
It should hold the variables of the program
that we wish to be visible to the C side,
behind IORefs (so that they can be modified
without having to create another AppState object).

For example, if we write a program having a counter
that we would like to present in a GUI,
we can add an IORef Integer.
Here is the example
also seen in Section \ref{subsubsec:evencounter-backend}.
(In Haskell, it becomes a \texttt{data},
but with basically the same description.)

\begin{lstlisting}[language=Haskell]
record AppState : Set where
  constructor MkAppState
  field
    counterRef : IORef Integer
open AppState public
{-# COMPILE AGDA2HS AppState #-}
\end{lstlisting}

Without the IORef wrappers, for every single change in the counter value,
we would need to create a new AppState instance,
then pass a new StablePtr to the C side,
while freeing the previous one.

We are going to continuously hold a StablePtr
to a single AppState object,
wrapped inside a C++ object of class HsAppStateWrapper,
and then pass this around throughout the entire lifetime of the program.
It is a typical example of RAII in C++:
it creates the StablePtr (as well as the object behind it) in the constructor
and frees it in the destructor.
Again, an example:

\begin{lstlisting}[language=C]
class HsAppStateWrapper {
  private:
    const HsStablePtr appStatePtr;

  public:
    HsAppStateWrapper():
      appStateWrapper(haskellFFICallCreatingAppState()) {}
  
    bool incrementWith(int toAdd) {
      someHaskellFFICall(appStatePtr, toAdd);
    }
    
    ~HsAppStateWrapper() {
      // a built-in FFI call
      hs_free_stable_ptr(appStatePtr);
    }
};
\end{lstlisting}

How this object itself is embedded in other classes, however,
is not trivial.
Containing a full HsAppStateWrapper in the view model itself
makes sense,
as it only contains a StablePtr,
which is technically a \texttt{void*}
(so it has the same size as a pointer).
The view, however, contains a reference
to the view model instance,
as that consists of several non-trivial objects.
We have considered hiding the HsAppStateWrapper behind a reference, too,
for the sake of consistence;
but we finally dropped the idea due to the obvious inefficiency.



\subsection{Using Haskell dependencies} 
\label{subsec:haskell-dependencies}

Many times throughout the code,
we need Haskell concepts that are unavailable
in either Agda itself or the agda2hs library,
usually because of their highly technical nature
for which Agda has not originally been intended
(e.g.\ MVars and StablePtrs for futures).
Here, we enumerate possible approaches
to tackle this situation,
which, to the best of our knowledge,
has not been done before.

As an example, we are going to use
a function initialising the backend
in Section \ref{subsubsec:evencounter-backend},
which calls the Haskell function \texttt{newIORef}
from Data.IORef.

The solutions considered:

\begin{itemize}
\item Simply omitting the Agda definition
  and writing all affected definitions
  in a \textbf{foreign pragma}
  (or even in separate Haskell files).
  For our example,
  this could be done like this:

\begin{lstlisting}[language=Haskell]
{-# FOREIGN AGDA2HS
initApp :: IO (StablePtr AppState)
initApp = newStablePtr =<< (MkAppState <$> newIORef 0)

-- and everything depending on this
-- has to be written to a foreign pragma as well
-- ...
#-}
\end{lstlisting}

  This is the most straightforward approach,
  and the one primarily used in \textbf{Acorn}
  (see Section \ref{subsec:acorn}).
  However, we would strongly discourage its use:
  \begin{itemize}
  \item It is infective: once we write a definition this way,
    we have to write every other definition in every module
    depending on it
    as foreign pragmas.
    Afterwards, we cannot write Agda proofs
    for any of these definitions.
  \item It is hard to explain and handle for newcomers,
    which makes it especially unfortunate
    for a general solution like a software development kit.
  \end{itemize}
\item Defining \textbf{the function depending on} the Haskell library
  \textbf{as an Agda postulate},
  then including the real implementation
  in a foreign pragma:

\begin{lstlisting}[language=Haskell]
-- this is going to be used in Agda
postulate
  initApp : IO (StablePtr AppState)
{-# COMPILE AGDA2HS initApp existing-class #-}
{-# FOREIGN AGDA2HS
initApp :: IO (StablePtr AppState)
initApp = newStablePtr =<< (MkAppState <$> newIORef 0)
#-}

-- and now, functions depending on initApp
-- can be written in Agda as well
\end{lstlisting}

  Note the existing-class pragma below the definition;
  it is needed because as of the current version of agda2hs
  (commit \texttt{acb521e} on the master branch),
  the compiler throws an error whenever it encounters a non-erased Agda definition
  without a COMPILE pragma.
  The existing-class pragma basically mutes this error,
  allowing us to proceed without a compiler-generated Haskell definition
  and define our own by hand.
  Probably, this is not the use case originally intended,
  as the pragma is only used in the agda2hs standard library
  and does not appear in any test case
  \citep{agda2hs};
  despite that, it is completely fit for our purposes.

  This improves much on the previous solution:
  \begin{itemize}
  \item It lets us stay in the Agda world for the rest of the code,
    provided we do not use Haskell dependencies again.
  \item It is easier to understand,
    as the type signature remains in Agda,
    with the ``magic'' staying in the background.
  \item It provides an ad-hoc solution quickly,
    with relatively little boilerplate code.
  \end{itemize}

  However, if we use the same dependency (newIORef) later again,
  we need foreign pragmas for every definition using it,
  having to write boilerplate code every time.

  For the above reasons, we consider this a convenient approach
  only if we only use a given Haskell dependency once.

\item Defining the required \textbf{Haskell elements themselves}
  as \textbf{Agda postulates},
  linking to the original Haskell import in a foreign pragma,
  then using the postulate
  in the function we want to define.
  Now, our example looks like this
  ($\equiv$ is written as \texttt{\textbackslash{}equiv} for technical reasons):

\begin{lstlisting}[language=Haskell]
{-# FOREIGN AGDA2HS
-- a Haskell import statement
-- (in the other solutions, this is not qualified)
import qualified Data.IORef
#-}

-- these are going to be used in Agda
postulate
  IORef : Set -> Set
  newIORef : {a : Set} -> a -> IO (IORef a)
  -- ...

  -- even some laws can be postulated here:
  @0 readNewIORef : x : a -> (readIORef =<< newIORef x)
                                     \equiv return x

-- in a foreign pragma, we define these
-- as synonyms for the original dependencies
{-# COMPILE AGDA2HS IORef existing-class #-}
{-# COMPILE AGDA2HS newIORef existing-class #-}
{-# FOREIGN AGDA2HS
type IORef a = Data.IORef.IORef a

newIORef :: a -> IO (IORef a)
newIORef = Data.IORef.newIORef
#-}

-- and the later usage can stay entirely in Agda:
initApp : IO (StablePtr AppState)
initApp = newStablePtr =<< (MkAppState <$> newIORef 0)
\end{lstlisting}

  We can also think of the postulate
  as an FFI call to Haskell,
  as the actual code is defined there
  and the Agda program just refers to that.

  \begin{itemize}
  \item With this approach,
    a Haskell dependency (in this case, newIORef)
    can be used over and over again
    with only a single foreign pragma.
  \item As in the example, some laws can also be postulated for the definitions,
    which gives us new opportunities for correctness proofs.
    \begin{itemize}  
    \item We can even choose to actually reimplement
      the referenced function in Agda
      instead of only postulating it,
      while compiling it to the fast Haskell built-in.
    \end{itemize}
  \end{itemize}

  This solution might be suitable for dependencies used frequently
  in a specific module of the project.

  The main drawback of the approach
  is the amount of boilerplate code it generates,
  especially as every single dependency used needs to be postulated.
  Therefore, for a single usage of a Haskell definition,
  it might unnecessarily complicate the code.
  There has been a discussion on
  including keywords in agda2hs
  meant specifically for this purpose:
  a concept called a
  ``COMPILE ... IMPORT'' pragma \citep{agda2hs-compile-import}.

\item Another mitigation of the boilerplate problem is
  moving such Haskell references into one or more separate Agda
  \textbf{modules with names starting with the prefix \texttt{Haskell.}}.
  In this example, IORef and newIORef from the Data.IORef module
  go into the module \texttt{Haskell.Data.IORef},
  and are as short as:

\begin{lstlisting}[language=Haskell]
postulate
  IORef : Set -> Set
  newIORef : {a : Set} -> a -> IO (IORef a)
  -- again, laws can also be postulated here
\end{lstlisting}

  Since pull request \href{https://github.com/agda/agda2hs/pull/379}{\#379},
  agda2hs treats such modules specially,
  producing no compiled Haskell files from them
  and automatically importing the original Haskell module instead
  in other modules.
  This can be done for modules imported from an arbitrary Hackage library as well,
  and in fact, this is what we do for our cfuture library
  (which adds its definitions to Control.Concurrent.CFuture).

  This might be the ideal approach
  for dependencies used many times,
  especially if they occur across modules
  (in which case, it is more transparent not to have it
  hidden inside one module
  that contains some of our own code as well).
  Still, for using a single Haskell dependency
  and that only a few times,
  it may be an overkill
  compared to the previous solution.

  The \textbf{EvenCounter} example project,
  written specifically for demonstration in this paper,
  mostly relies on this solution.
  (See Section \ref{subsec:evencounter}.)

\item If the \textbf{dependency} is frequent enough
  to be useful for any agda2hs-based project,
  it can be \textbf{added as a postulate
  to the agda2hs standard library}
  (or even reimplemented there in Agda).

  Before PR 379,
  this was the only way to have Haskell dependencies
  translated automatically to Haskell imports;
  as we have added a significant amount of code to the library
  that is used by no one else, however,
  we have diverged significantly
  from the vanilla agda2hs distribution
  (for a time, we even had to use a \textbf{custom branch},
  \href{https://github.com/viktorcsimma/agda2hs/tree/the-agda-sdk}{the-agda-sdk},
  for our projects).
  However, this solution became obsolete with the new feature,
  and it enabled us to return to the vanilla agda2hs compiler.

  Thus, adding definitions to the agda2hs standard library
  is now recommended only if they have a significant value
  to other users of the compiler as well.

\item As an alternative to the approaches before,
  agda2hs has a feature
  called \textbf{rewrite rules},
  through which one can provide Haskell concepts
  to be substituted for certain Agda definitions,
  via a YAML configuration file.
  It has originally been designed
  by the author of this paper \citep{rewrite-rules-pr}.
  Here is an example,
  mapping some arithmetic functions for the Rational type of
  the original Agda (\emph{not} agda2hs!) standard library
  to functions in Haskell Prelude:

\begin{lstlisting}[language={}]
rules:
  - from: Data.Rational.Unnormalised.Base._+_
    to: _+_
    importing: Prelude
  - from: Data.Rational.Unnormalised.Base._*_
    to: _*_
    importing: Prelude
  - from: Data.Rational.Unnormalised.Base.-_
    to: negate
    importing: Prelude
  - from: Data.Rational.Unnormalised.Base._-_
    to: _-_
    importing: Prelude
\end{lstlisting}

  According to the documentation \citep{agda2hs-docs},
  this is primarily meant for projects

\vspace{2mm}

  \emph{``depending on a large library which is not agda2hs-compatible (e.g the standard library). In this case, you might not want to rewrite the entire library, but may still rely on it for proofs.''}

\vspace{2mm}
  In the example, this large library is the aforementioned
  Agda standard library,
  which has not been designed with agda2hs in mind,
  and would otherwise be unusable here.

  For using simple Haskell definitions,
  rewrite rules have been argued to be inefficient,
  as they provide no type-checking
  and need duplicate information
  separate from the module itself \citep{agda2hs-compile-import}.
  Actually, in the example of newIORef,
  this would mean postulating the dependency
  \emph{and} adding a rule to the YAML file
  (instead of doing everything within the code itself),
  which could perhaps move some boilerplate
  out from the main files,
  but would actually make the source code
  harder to understand.

  Neither of the projects presented here
  rely on this feature.
\end{itemize}

\subsection{Futures}
\label{subsec:futures}

The concept of futures is named after the \texttt{std::future} class
of the C++ standard library \citep[Section 33.10]{cpp23-draft}.
Similarly to that, a future represents a calculation
which has not produced a result yet,
enabling the holder to wait for it or detach it.

Futures are an almost 50-year-old concept---%
the term was coined by \citet{future-1977}---%
and there have already been implementations of futures in Haskell as well;
e.g.\ that of \citet{old-future}.
What makes our solution unique
is the ability to arbitrarily interrupt the calculation
(either from inside Haskell or from C/C++),
as well as the ease of exporting Haskell futures to C/C++
at all.

As this implementation might be useful in other Haskell projects
independently of our use case,
the implementation has been uploaded to Hackage
under the name \emph{\href{https://hackage.haskell.org/package/cfuture}{cfuture}}.
Agdalache provides an Agda wrapper to this library,
using \texttt{Haskell.}-prefixed modules.

\subsubsection{Motivation}
\label{subsubsec:future-motivation}

In any GUI application,
if we have to include a (possibly) long calculation,
which forces the user to wait,
we need to run it on a separate thread.
Otherwise, if the process runs on the UI thread,
the UI itself becomes unresponsive,
which appears to the outside as a freeze
\citep{android-threads}.
Also, the user probably expects a way
to interrupt the calculation,
in case they decide not to wait for it any more.

To make this possible
while also keeping the API to the Haskell backend simple,
we have considered the following options.

\begin{itemize}
\item \textbf{Simple (synchronous) calls to Haskell functions exported to C}
  (the original approach for which the GHC FFI has been designed).
  These are easy to understand, especially for backend developers.
  However, these block the calling thread entirely,
  with no option for safe interruption whatsoever
  \citep[FFI]{ghc-docs-ffi}.
\item \textbf{Synchronous calls to Haskell functions
  wrapped into a \texttt{C++ std::thread}.}
  This makes it possible to run other actions while waiting for the result
  and even to run some actions (``triggers'') on completion.
  The main drawback of this approach is that
  it does not provide for interrupting the calculation,
  as such threads cannot be interrupted safely.
  The problem cannot be circumvented
  with an \texttt{std::jthread} either:
  the code inside has to manually check for a stop request
  \citep[Section 33.4.4]{cpp23-draft},
  which it cannot do during a Haskell foreign call.
  In other words, there is no way to safely interrupt a Haskell foreign call
  without providing some mechanism in the backend itself.
\item Utilising a so-called \textbf{``watcher thread''} on the \textbf{Haskell side}
  which, when woken up, interrupts the calculation thread
  with a \texttt{throwTo} call.
  A further question is how this watcher thread can be activated
  from the frontend,
  for which we present two approaches:
  \begin{itemize}
  \item Using an \textbf{OS semaphore}
    (POSIX semaphores on Linux or Win32 events on Windows).
    Here is an example for POSIX:
    
\begin{lstlisting}[language=Haskell]
runInterruptibly :: IO a -> a -> IO a
runInterruptibly action resultOnInterrupt = do
  (mVar :: MVar (Maybe a)) <- newEmptyMVar

  -- runs the calculation itself
  childThreadId <- forkIO
                     (putMVar mVar =<< (Just <$> action))

  -- here, the name is simply a fixed string
  semaphore <- semOpen "AcornInterruptSemaphore"
                       (OpenSemFlags True False)
                       stdFileMode
                       0

  watcherThreadId <- forkIO $ do
    ... -- wait for the semaphore;
        -- then if triggered, kill the childThread 
        -- and write Nothing to the MVar 

  maybeResult <- readMVar mVar
  case maybeResult of
    Just result -> do
      -- there is a result!
      -- we make the watcher thread end gracefully
      -- by unlocking the semaphore
      semPost semaphore
      return result
    Nothing -> return resultOnInterrupt
      -- this means that the watcher thread
      -- has already interrupted the calculation
\end{lstlisting}
    
    This is very convenient if the program can only have
    one background calculation at the same time,
    like in \textbf{AcornCalc},
    as we can simply use a fixed semaphore name
    (here ``AcornInterruptSemaphore'').
    If there are more than one, however,
    the developer has to think about how they could use
    different semaphore names for different calculations---%
    in the worst case, they will have to create a new name for every call,
    which makes this approach drastically more complex.
    (Much better is when one can create a finite number of categories
    with at most one calculation each---%
    then, every category can be given a single name.)
    Also, the code has to be written separately for each platform,
    and the implementation is error-prone
    (with open semaphores left behind in the system,
    the program might malfunction on the next run).

    While semaphores had been useful for AcornCalc,
    we cannot choose them as a general solution
    due to the aforementioned limitations.
  \item Using \textbf{PrimMVars and the \texttt{hs\_try\_putmvar} function},
    as the documentation \citep[FFI]{ghc-docs-ffi} suggests.
    Since this is the ``official'' method of waking up a Haskell thread,
    it is not a new idea by itself.
    However, we can create an object
    tied specifically to the calculation
    and able to provide its result as well,
    by bundling
    an interruption MVar and a calculation MVar
    in a single object,
    which can also be exported to C via a pointer---%
basically, we get a \textbf{future} object,
    exportable to C/C++
    and interruptible anytime.
    What is more, on the C side,
    interrupting by \texttt{hs\_try\_putmvar} is much faster
    than a full FFI call.

    After modelling some Haskell dependencies with postulates,
    the code can mostly be re-implemented in Agda as well,
    as a \texttt{Haskell.}-prefixed module.
    (For details,
    see also Section \ref{subsec:haskell-dependencies}.)

    The drawback is that due to the type variable the MVar contains,
    instantiations for different types are needed to retrieve data from the futures,
    which becomes especially tiresome when writing a C++ wrapper class.
    This means many boilerplate functions for different C primitive types,
    which largely do the same thing.
    (From these, only the ones
    for \texttt{int}, \texttt{void} and pointers have been written so far,
    but others can be added similarly.)

    Also, futures do not, by themselves, allow triggers
    (i.e.\ code executed automatically on completion);
    this again needs \texttt{std::thread} instances,
    and a correct implementation is not trivial.
    An experimental extension is discussed
    in Section \ref{subsubsec:triggerfuture}.
  \end{itemize}
\end{itemize}

Agdalache mainly uses \textbf{PrimMVars}.
The semaphore-based solution is provided as a legacy option as well.

\subsubsection{Overview}

To the best of our knowledge, our design of Haskell futures
is both the first to allow arbitrary interruption
and the first that can be exported to C/C++ via the FFI
(where it can also be arbitrarily interrupted).
This means a significant advancement
compared to previous implementations
like \citep{old-future}.

While the design pattern is more-or-less reproducible
in other languages as well,
we are going to present it via
the concrete Agda/Haskell-specific implementation,
as it is easier to understand this way.

On the Agda/Haskell side, a future (\texttt{Future a}) is an algebraic data type
containing two MVars:
an \texttt{MVar ()} for interruption
and an \texttt{MVar (Maybe a)} for the result.

\begin{lstlisting}[language=Haskell]
data Future a = MkFuture 
  (MVar ())
  (MVar (Maybe a))
\end{lstlisting}

This can be exported to C
by writing StablePtrs for the two MVars
to a memory location (a pointer) provided by the caller.
Thus, a CFuture on the C side is an array of two StablePtrs:

\begin{lstlisting}[language=C]
typedef HsStablePtr CFuture[2];
\end{lstlisting}

For creating futures from given calculations
(that is, \texttt{IO a} actions),
 \texttt{forkFuture} and \texttt{forkFutureC} can be used,
with the latter also writing the two StablePtrs
to a location given as a C pointer.

An example for a Haskell function starting a calculation
and exporting a corresponding Future to C:

\begin{lstlisting}[language=Haskell]
longBackgroundCalculation :: Int -> IO Int
longBackgroundCalculation n = ...

-- CFuturePtr is a pointer to the C array
longBackgroundCalculationC :: Int -> CFuturePtr -> IO ()
longBackgroundCalculationC n ptr
                = forkFutureC ptr (longBackgroundCalculation n)

foreign export ccall longBackgroundCalculationC
  :: Int -> CFuturePtr -> IO ()
\end{lstlisting}

And in C headers, it will be noted like this
(note that technically, \texttt{CFuture} is now a \texttt{HsStablePtr*}):

\begin{lstlisting}[language=C]
extern void longBackgroundCalculationC
  (int someParameter, CFuture future);
\end{lstlisting}

On the C side, we need different functions
for all different C primitive types,
like \texttt{getC\_Int} for integers,
\texttt{getC\_Ptr} for pointers etc.\
(this is the source of the problems mentioned
in Section \ref{subsubsec:future-motivation}).
Interruption occurs by calling \texttt{hs\_try\_putmvar}
on the first StablePtr---%
on interruption, the watcher thread writes
\texttt{Nothing} to the result MVar,
and afterwards, \texttt{getC} returns false.
\texttt{hs\_try\_putmvar} also frees the StablePtr given;
in other cases, we have to manually keep track of the StablePtrs,
as they are not freed automatically.

The following example demonstrates how it is possible (however inconvenient)
to handle futures using only C and the Haskell FFI.

\begin{lstlisting}[language=C]
CFuture future; /* HsStablePtr future[2]; */
/* starting calculation and creating the MVars themselves: */
longBackgroundCalculationC(42, future);

/* ... */

/* then, either we obtain the result and free the pointers: */
int result = getCIntFromFutureC(future);
hs_free_stable_ptr(future[0]);
hs_free_stable_ptr(future[1]);

/* or, we interrupt via the interruption MVar
   (which thereby gets freed automatically)
   _and_ free the other one */
hs_try_putmvar(future[0]);
hs_free_stable_ptr(future[1]);

/* failure to free causes a memory leak */
\end{lstlisting}

With an RAII-style C++ wrapper class
(also called \texttt{Future}),
handling these obstacles becomes much easier and safer.
The constructor takes an \texttt{std::function<void(CFuture)>} \citep[Section 22.10.17]{cpp23-draft},
which should essentially be a Haskell callback
starting the calculations and writing the StablePtrs to a given location.
These can easily be constructed with lambdas
calling backend functions (exported via \texttt{forkFutureC})
with the pointer and captured parameters.
From then on, handling the future will be easy,
with various flags and simple getter methods available,
as well as an \texttt{interrupt()} call.

The destructor checks whether the future has
either been queried or interrupted,
and if not, it automatically interrupts the calculation
(like \texttt{std::jthread} does).

Now, the above example becomes much more concise:

\begin{lstlisting}[language=C]
// we need a lambda that takes a HsStablePtr* and returns void
Future<int> future(
  [](CFuture futurePtr){
    longBackgroundCalculationC(42, futurePtr);
  }
);

// ...

// either we obtain the result and free the pointers:
int result = future.get();

// or we interrupt:
future.interrupt();

// pointers cannot leak,
// as the destructor gets called at the end of the function
// (this also interrupts the calculation, if it is still running)
\end{lstlisting}

\subsubsection{Running triggers}
\label{subsubsec:triggerfuture}

An experimental feature is
the \texttt{TriggerFuture} C++ class,
a subclass of \texttt{Future}:
this cannot be waited for,
but instead executes triggers provided as \texttt{std::function} objects
on completion.
It actually waits for the underlying future
and then executes triggers
on an \texttt{std::thread},
while still providing the possibility of interruption.

An important problem to solve still remains:
the chaining of futures in general.
The existing object cannot be replaced with a new one
in a trigger,
because by deleting the future, the very function object being executed
would be destroyed as well.

In EvenCounter (described in Section \ref{subsec:evencounter}),
the ad-hoc solution is
to run normal futures one-after-one in a for-loop
on an \texttt{std::thread};
freeing them immediately on termination
and replacing them with the next one.
See \href{https://github.com/viktorcsimma/even-counter/blob/master/frontend/src/ViewModel/MainViewModel.cpp}{\texttt{MainViewModel.cpp}}.

\section{The toolkit}

Here, only a quick overview
of the structure and usage of the development kit
is given.
For tutorials and implementation details,
we again refer to Section \ref{sec:repos-and-links},
with the documentation and the code.

\subsection{Separate compilation of backend and frontend}

The default structure of the project skeleton
actually contains three fully functional CMake projects:
one for the backend, one for the frontend
and a root project embedding both.
This enables the developer to compile
only the backend (when they do not need a GUI frontend),
only the frontend (e.g.\ for a globally installed backend library),
or both together.

\subsection{Backend}

Compilation of the backend technically happens
via:
\begin{itemize}
\item calling agda2hs on a file called \texttt{All.agda},
  containing references to Agda and Haskell files to be included;
\item then calling GHC on the previously generated \texttt{All.hs} file,
  or on \texttt{Main.hs} for a command-line binary (or on \texttt{TestMain.hs} for QuickCheck tests).
\end{itemize}

In the first step, the Agda type checker built into agda2hs
verifies Agda source code,
and its agda2hs backend translates those definitions needed,
along with foreign pragmas,
into machine-generated (but human-readable)
Haskell modules.
GHC then compiles these
into a static library
or an executable.

A disadvantage is that
\texttt{All.agda} has to be continuously updated:
files to be checked by the type checker
are imported in the Agda part of the file,
while those to be compiled by GHC
should be written into the foreign Haskell part.
Automatic generation of the file might be
a possible future improvement,
similarly to the \texttt{Everything.agda} file of
the Agda standard library
\citep[\texttt{GenerateEverything.hs}]{agda-stdlib}.

Communication with the frontend is done
via the Haskell foreign function interface \citep{ghc-docs-ffi}.
Exported functions can be in any source file
(the skeleton and example projects
collect them under \texttt{Interaction.agda},
for the sake of consistency);
the corresponding C type signatures
should be copied into headers
in the \texttt{include} folder of the backend.

Pre-written tools can be found
under \texttt{src/Tool} of the backend part.
These include
some data structures tailored for agda2hs usage,
definitions for usage in proofs,
postulates for foreign C types
and the backend Agda wrapper of futures.

\subsection{Frontend}

The frontend is, by default,
a Qt-based C++ GUI application
(but technically, any C++ or C code
can be attached to the backend).
The tools provided in the skeleton project
include wrappers for backend concepts
(including futures),
as well as a default structure
for a view model and a view.

\subsection{Testing}

It is often useful to already have some tests at hand,
especially if one has not yet begun writing a formal verification for the program.
Also, a test can capture certain aspects
that are usually not covered by a formalisation;
e.g.\ the correctness of UI elements
or integration into a broader software environment.

Agdalache provides three mechanisms
through which one can write test cases quickly:

\begin{itemize}
\item \textbf{Agda tests}, evaluated by the type checker
  before compilation,
  can actually be placed anywhere in the backend,
  but the recommended location is under the Test folder.
  An example is defining a simple equality
  between two concrete values
  (e.g.\ a function call and its result)
  with \texttt{refl}:\\
\texttt{
test2 : eitherAddInteger 0 0 $\equiv$ Right 0\\
test2 = refl\\
test3 : eitherAddInteger 1 0 $\equiv$ Left "first parameter is odd"\\
test3 = refl
}

  These can later be expanded
  to full-fledged proofs.
\item \textbf{Haskell QuickCheck} tests,
  run by issuing \texttt{cabal test},
  can be written under Test/Haskell in the backend.
  These are based on Boolean predicates
  that are then tested against
  a huge amount of randomly generated inputs.
  They can even be written in Agda,
  besides some technical code
  in a foreign Haskell block at the end of the file.

  For the previous function,
  an example looks like this:
\begin{lstlisting}[language=Haskell]
prop_correctWithTwoEven : Integer -> Integer -> Bool
prop_correctWithTwoEven x y =
  eitherAddInteger (2 * x) (2 * y)
    == Right (2 * x + 2 * y)
{-# COMPILE AGDA2HS prop_correctWithTwoEven #-}

-- Similarly when one of the parameters is not even.
prop_correctWithOddAndEven : Integer -> Integer -> Bool
prop_correctWithOddAndEven x y =
  eitherAddInteger (2 * x + 1) (2 * y)
    == Left "first parameter is odd"
{-# COMPILE AGDA2HS prop_correctWithOddAndEven #-}
\end{lstlisting}
\item Ordinary \textbf{Catch2}-based \textbf{C++} test cases
  can be added under \texttt{src/Test}
  in the frontend.
  As Catch2 is a well-established test library
  used unaltered in Agdalache,
  we simply refer to that project's documentation
  \citep{catch2}.
\end{itemize}

\section{Benchmarks}
\label{sec:benchmarks}

One might well ask:
does the proposed methodology stand on equal ground
with current solutions?
In this section, we are going to show
that besides utilising the special properties of Agda,
agda2hs produces binaries with a reasonable speed;
hence, Agda developers can use it
without worrying about performance penalties.

Including these benchmarks in this paper might be confusing,
as the rest of the article is very specifically about
solutions \emph{utilising} agda2hs,
while the benchmark concerns agda2hs itself.
However, if the compiler turned out to be prohibitively inefficient
compared to existing competitors,
the entire project would be of little value.
We demonstrate here that this is not the case.

\subsection{Hardware configuration}

Benchmarks have been done on an Ubuntu 24.04 machine
with the following specifications:

\begin{itemize}
\item CPU: AMD Ryzen 7 3700U;
\item GPU: Radeon RX Vega 10;
\item system memory: 16 GiB;
\item storage: SK Hynix 256 GB NVMe-M.2 SSD.
\end{itemize}

\subsection{Methodology}

The source files are based on code which
András \citet{smalltt} has previously used
for similar benchmarks;
more precisely, \texttt{conv\_eval.agda} and \texttt{conv\_eval.v}.
There, he defines a lambda-based representation
of natural numbers and binary trees,
then forces the type checker to decide on
the equality of certain complex expressions,
or to evaluate them.
This makes it possible to measure
the efficiency of various type checkers.

Compared to the original \emph{smalltt} source code,
definitions are rewritten
with Haskell-style type classes,
thereby circumventing the need for
\texttt{--type-in-type}
and enabling translation to Haskell
(but worsening code readability).
For compilation tests, we add code
which evaluates expressions
on an extracted function;
we then compile and run that function
to measure the time needed
for:
\begin{itemize}
\item \textbf{translation},
\item \textbf{compilation}, and
\item \textbf{running}.
\end{itemize}

For time measurements,
we use the built-in \texttt{time} command of Bash
(instead of the \texttt{time} utility of Unix,
as the latter is usually less precise).

Before each measurement, we comment out
every unnecessary definition
and check or compile only the one needed---%
surprisingly, it often makes elaboration
several seconds faster
if previous, less complex objects
have been checked before.

On the main function (which is to be compiled to an executable),
we follow a structure similar to
the \texttt{conv\_eval.agda} and \texttt{conv\_eval.v} files
of \citet{smalltt},
by evaluating either:
\begin{itemize}
\item a natural number X to a machine integer (referred to as \emph{nX}, with a number substituted for X); or
\item a full binary tree of depth X, with True values in the leaves,
  to the Boolean AND of the values in the leaves;
  with \texttt{forceTree} (similarly referred to as \emph{tX}).
\end{itemize}

For OCaml, a main function has to be added by hand
in order to get the source code compiled;
see \texttt{prepare\_ocaml.sh}.

\subsection{Results}
\label{subsec:results}

The question is whether agda2hs really is competitive
compared to existing solutions:
the default GHC backend of Agda (MAlonzo)
and the OCaml extraction tool of Rocq.
The type-checking performance of Agda and Rocq had already been compared
by \citet{smalltt},
and as it has previously been shown,
Rocq is definitely superior to Agda in these fields.
However, when evaluating agda2hs, it is more interesting how the various backends have performed.

All the measurements in the tables are in seconds;
smaller values are better.
Where three values appear for a given compiler,
the measurements are for the
\texttt{-O0}, \texttt{-O1} and \texttt{-O2} flags
of GHC, respectively.

\subsubsection{Translation to the implementation language}

\begin{table}[h]
\centering
\caption{Translation times (sec)}
\begin{tabular}{rccc}

    \hline
    ~ & \multicolumn{1}{c}{Agda MAlonzo} & \multicolumn{1}{c}{agda2hs} & \multicolumn{1}{c}{Rocq OCaml extraction} \\
    \hline
    n2 & 1.162 & 0.610 & 0.162 \\
    n100 & 1.172 & 0.590 & 0.156 \\
    n10k & 1.173 & 0.600 & 0.140 \\
    n100k & 1.195 & 0.559 & 0.152 \\
    n1M & 1.163 & 0.582 & 0.151 \\
    n5M & 1.133 & 0.581 & 0.146 \\
    n10M & 1.193 & 0.560 & 0.153 \\
    n100M & 1.162 & 0.572 & 0.144 \\
    n500M & 1.807 & 0.955 & 0.153 \\
    n1G & 1.837 & 0.976 & 0.149 \\
    t15 & 1.101 & 0.621 & 0.164 \\
    t18 & 1.090 & 0.613 & 0.148 \\
    t19 & 1.142 & 0.615 & 0.149 \\
    t20 & 1.081 & 0.592 & 0.143 \\
    t21 & 1.143 & 0.603 & 0.145 \\
    t22 & 1.091 & 0.602 & 0.139 \\
    t23 & 1.131 & 0.602 & 0.145 \\
    \hline

\end{tabular}
\label{tab:translation}
\end{table}

\begin{figure}[h]
  \centering
  \caption{Translation times for natural number evaluations. Note the logarithmic scale. \label{fig:translation-n}}
    \footnotesize
    \includesvg[width=0.8\textwidth]{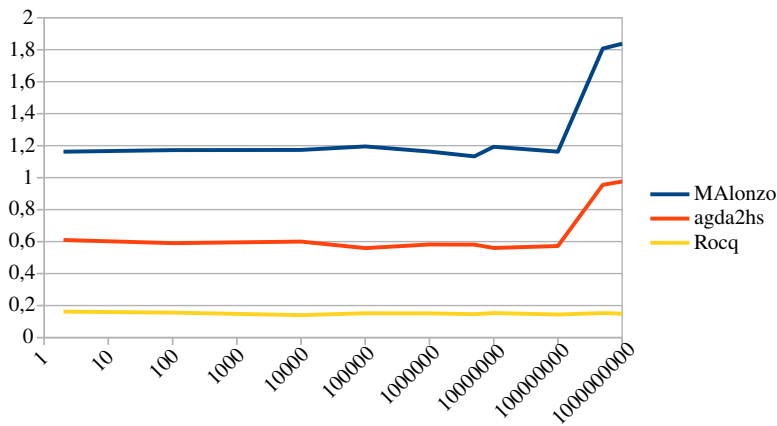}
\end{figure}

\begin{figure}[h]
  \centering
  \caption{Translation times for binary tree evaluations. \label{fig:translation-t}}
    \footnotesize
    \includesvg[width=0.8\textwidth]{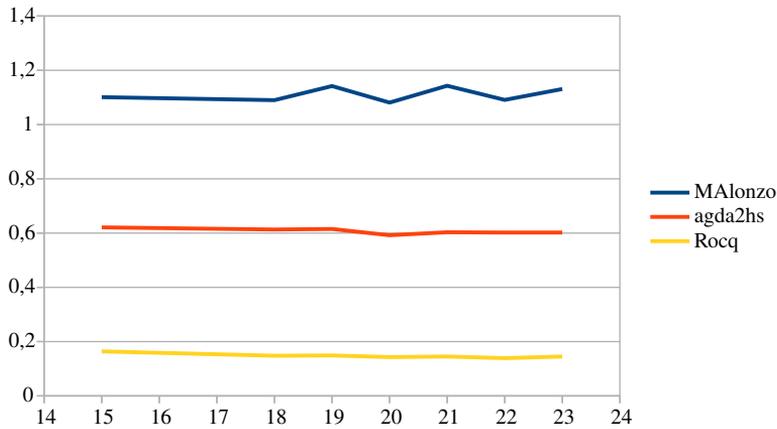}
\end{figure}

The results are presented in Table \ref{tab:translation}
and Figures \ref{fig:translation-n} and \ref{fig:translation-t}.
As expected,
translation times seem to be dependent on code structure,
rather than the concrete values used in the examples.
agda2hs is about twice as fast as MAlonzo is;
one of the reasons might be
that agda2hs just imports Haskell libraries,
while MAlonzo has to refactor them before importing
(for they are by themselves uncompatible with the Haskell code
generated from Agda files).
Rocq OCaml extraction is by far the fastest at this phase,
with translation times about four-six times lower
than those of agda2hs.

\subsubsection{Compilation to executables}
\label{subsubsec:executables-benchmark}

\begin{table}[h]
\centering
\caption{Compilation times (sec)}
\begin{tabular}{rccccccc}

    \hline
    ~ & \multicolumn{3}{c}{Agda MAlonzo} & \multicolumn{3}{c}{agda2hs} & \multicolumn{1}{c}{Rocq OCaml extraction} \\
    ~ & \multicolumn{1}{c}{\texttt{-O0}} & \multicolumn{1}{c}{\texttt{-O1}} & \multicolumn{1}{c}{\texttt{-O2}}
    & \multicolumn{1}{c}{\texttt{-O0}} & \multicolumn{1}{c}{\texttt{-O1}} & \multicolumn{1}{c}{\texttt{-O2}}
    & ~\\
    \hline
    n2 & 1.303 & 1.819 & 1.823 & 0.915 & 0.934 & 0.950 & 0.017 \\
    n100 & 1.306 & 1.791 & 1.900 & 0.850 & 0.901 & 0.929 & 0.021 \\
    n10k & 1.351 & 1.743 & 1.847 & 0.885 & 0.929 & 0.910 & 0.016 \\
    n100k & 1.334 & 1.787 & 1.818 & 0.868 & 0.914 & 0.924 & 0.022 \\
    n1M & 1.291 & 1.823 & 1.853 & 0.889 & 0.906 & 0.889 & 0.019 \\
    n5M & 1.339 & 1.755 & 1.829 & 0.871 & 0.911 & 0.902 & 0.024 \\
    n10M & 1.279 & 1.767 & 1.866 & 0.848 & 0.931 & 0.887 & 0.023 \\
    n100M & 1.334 & 1.805 & 1.869 & 0.878 & 0.886 & 0.925 & 0.018 \\
    n500M & 2.002 & 2.775 & 2.824 & 1.335 & 1.488 & 1.501 & 0.029 \\
    n1G & 1.985 & 2.752 & 2.886 & 1.393 & 1.434 & 1.449 & 0.017 \\
    t15 & 1.294 & 1.820 & 1.848 & 0.881 & 0.971 & 0.941 & 0.022 \\
    t18 & 1.308 & 1.807 & 1.826 & 0.890 & 0.925 & 0.948 & 0.021 \\
    t19 & 1.270 & 1.799 & 1.857 & 0.912 & 0.923 & 0.925 & 0.021 \\
    t20 & 1.268 & 1.806 & 1.826 & 0.872 & 0.913 & 0.917 & 0.022 \\
    t21 & 1.326 & 1.755 & 1.865 & 0.882 & 0.952 & 0.941 & 0.020 \\
    t22 & 1.312 & 1.796 & 1.856 & 0.866 & 0.922 & 0.927 & 0.017 \\
    t23 & 1.352 & 1.737 & 1.870 & 0.888 & 0.939 & 0.922 & 0.018 \\
    \hline

\end{tabular}
\label{tab:compilation}
\end{table}

\begin{figure}[h]
  \centering
  \caption{Compilation times for natural number evaluations. Note the logarithmic scale. \label{fig:compilation-n}}
    \footnotesize
    \includesvg[width=0.8\textwidth]{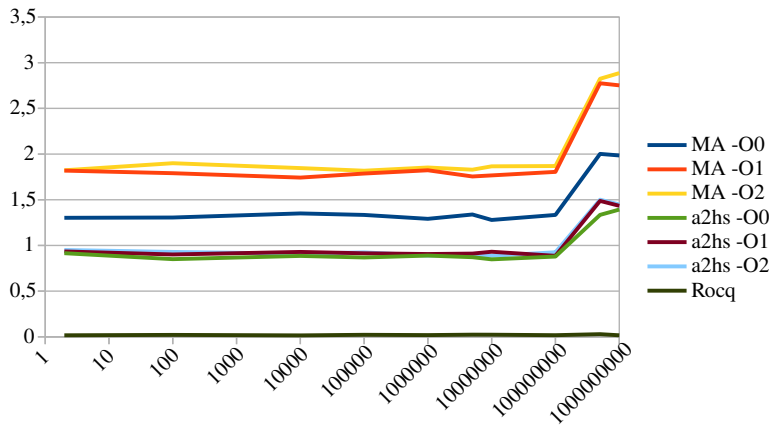}
\end{figure}

\begin{figure}[h]
  \centering
  \caption{Compilation times for binary tree evaluations. \label{fig:compilation-t}}
    \footnotesize
    \includesvg[width=0.8\textwidth]{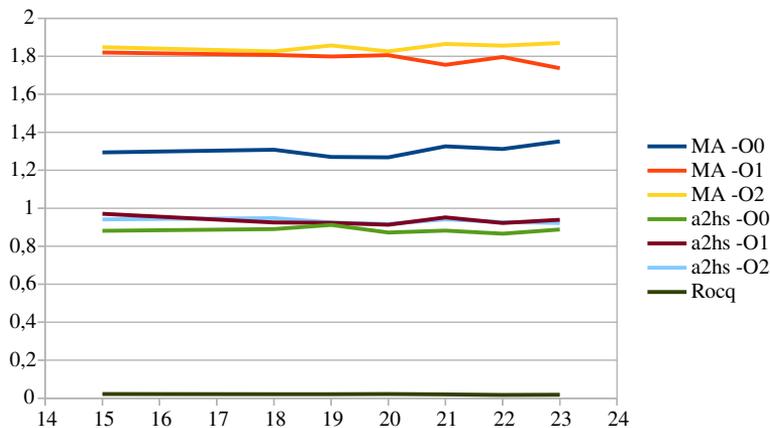}
\end{figure}

For the results, see Table \ref{tab:compilation}
as well as the corresponding Figures
\ref{fig:compilation-n} and \ref{fig:compilation-t}.
Running GHC on agda2hs-generated code
is again significantly faster
than on the type coercion-based format
written by MAlonzo
(although again, the reasons might also include
the larger number of files).

However, the OCaml compiler runs almost instantly
on the source code extracted from the Rocq module.
(We note that for Rocq,
the extraction phase already outputs only the definitions
that are really going to be used during computation,
thus taking some overhead off the compiler.
Still, that is probably not the only reason
for such a great difference.)

For MAlonzo, GHC optimisation flags add an overhead of around 30--40\%;
in the case of agda2hs, this is much less significant.

\subsubsection{Running the executables}

\begin{table}[h]
\centering
\caption{Runtimes (sec)}
\begin{tabular}{rccccccc}

    \hline
    ~ & \multicolumn{3}{c}{Agda MAlonzo} & \multicolumn{3}{c}{agda2hs} & \multicolumn{1}{c}{Rocq OCaml extraction} \\
    ~ & \multicolumn{1}{c}{\texttt{-O0}} & \multicolumn{1}{c}{\texttt{-O1}} & \multicolumn{1}{c}{\texttt{-O2}}
    & \multicolumn{1}{c}{\texttt{-O0}} & \multicolumn{1}{c}{\texttt{-O1}} & \multicolumn{1}{c}{\texttt{-O2}}
    & ~\\
    \hline
        n2 & 0.014 & 0.014 & 0.014 & 0.014 & 0.014 & 0.013 & 0.003 \\
        n100 & 0.013 & 0.013 & 0.014 & 0.013 & 0.013 & 0.012 & 0.006 \\
        n10k & 0.012 & 0.014 & 0.012 & 0.014 & 0.012 & 0.014 & 0.003 \\
        n100k & 0.023 & 0.024 & 0.013 & 0.034 & 0.013 & 0.014 & 0.004 \\
        n1M & 0.232 & 0.160 & 0.185 & 0.224 & 0.141 & 0.150 & 0.029 \\
        n5M & 0.951 & 0.832 & 0.813 & 0.766 & 0.191 & 0.181 & 0.073 \\
        n10M & 2.039 & 1.656 & 1.670 & 1.560 & 1.410 & 1.388 & 0.156 \\
        n100M & 21.008 & 18.686 & 18.721 & 15.242 & 2.290 & 2.291 & 1.502 \\
        n500M & ME & ME & ME & ME & ME & ME & 7.786 \\
        n1G & ME & ME & ME & ME & ME & ME & 15.224 \\
        t15 & 0.013 & 0.024 & 0.024 & 0.013 & 0.013 & 0.024 & 0.009 \\
        t18 & 0.054 & 0.033 & 0.044 & 0.034 & 0.033 & 0.035 & 0.049 \\
        t19 & 0.095 & 0.063 & 0.075 & 0.044 & 0.043 & 0.054 & 0.108 \\
        t20 & 0.184 & 0.124 & 0.113 & 0.063 & 0.074 & 0.083 & 0.188 \\
        t21 & 0.314 & 0.224 & 0.244 & 0.134 & 0.124 & 0.145 & 0.344 \\
        t22 & 0.604 & 0.423 & 0.404 & 0.255 & 0.244 & 0.234 & 0.709 \\
        t23 & 1.153 & 0.794 & 0.793 & 0.444 & 0.434 & 0.423 & 1.373 \\
    \hline
\end{tabular}
\label{tab:running}
\end{table}

\begin{figure}[h]
  \centering
  \caption{Runtimes for natural number evaluations.\\Note the logarithmic scale, as well as the memory exhaustion of Agda solutions.\label{fig:runtime-n}}
    \footnotesize
    \includesvg[width=0.8\textwidth]{charts/runtime_n.svg}
\end{figure}

\begin{figure}[h]
  \centering
  \caption{Runtimes for binary tree evaluations. \label{fig:runtime-t}}
    \footnotesize
    \includesvg[width=0.8\textwidth]{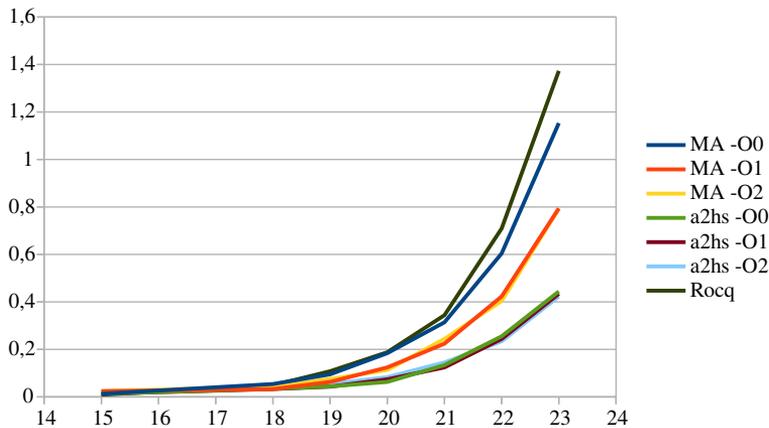}
\end{figure}

The final result table is Table \ref{tab:running},
further illustrated by Figures
\ref{fig:runtime-n} and \ref{fig:runtime-t}.
Here, ME means the program exhausted the system memory
and was killed by the OS.

As it can be seen, there is again a huge difference between MAlonzo and agda2hs.
Optimisation mostly helps agda2hs in the natural number tests
but MAlonzo in the binary tree tests;
we could not get to a conclusion on what might cause this.

Natural number extraction was really hard for Agda programs;
Rocq's advantage there is enormous.
Tree evaluation, however, is a point
where agda2hs-compiled executables even beat
those extracted from Rocq.

We have been thinking about an explanation for the difference.
One possible reason could be the lazy evaluation of Haskell
compared to the strict evaluation of OCaml;
but that does not answer
why it was the other way around
for natural number iteration.
This could be an interesting topic
for further research,
along with the effect of GHC optimisation flags
on MAlonzo and agda2hs.

\subsection{Threats to validity}

Multiple, mostly external, threats to the validity of the results
have been considered.

\begin{itemize}
\item Measurements have been performed only
for two very specific dummy tasks.
It might not be possible to generalise our findings
to a real-world project,
which would essentially require implementing a life-size project twice,
both in Agda and in Rocq.
\item Even when switching between these two tasks,
there is a visible difference in
the performance of even a single alternative,
which we could not explain yet.
This means it is even harder to predict
how the platforms would perform a real-world task.
\item Also, the only test machine was a personal computer,
which leaves open the possibility of different runtimes
when provided a server-grade size of RAM.
\end{itemize}
 
Still, the results clearly demonstrate that
there is no difference in order of magnitude
in the runtime of agda2hs-based versus that of Rocq-based programs,
which was our original goal with the benchmarks. 

\subsection{Conclusion}

As we have mentioned,
Rocq beats Agda when it comes to
type-checking and interactive calculations.
But for evaluating the capabilities of agda2hs and Agdalache,
the focus is on compilation time and runtime.

The OCaml extraction feature of Rocq is by orders of magnitude faster
than agda2hs
in both translation and compilation.
At run-time, results were not so conclusive,
with multipliers being much smaller.
For tree evaluation, Agda even beats Rocq
with both backends;
this might be attributed to a better handling
of parallelisation.
It can also be seen
that the \texttt{-O1} flag of GHC
often helped accelerate executables significantly.

It might seem as if
the superiority of Rocq at almost every field
made Agda obsolete.
Note, however,
that Agda has several advantages compared to Rocq;
namely, a cleaner syntax designed more for writing programs
rather than proofs
and a unique, helpful text editor mode.
With test results not being worse \emph{by orders of magnitude},
this might justify someone preferring Agda.
And for those preferring Agda,
agda2hs (and by extension, Agdalache)
might be an excellent choice.

Finally, agda2hs outperformed
the built-in GHC backend of Agda
at practically every field;
although we have to note that
they serve a different purpose:
MAlonzo is for compiling arbitrary Agda code to Haskell,
while agda2hs only supports
a Haskell-compatible subset of Agda.
Having said that, most practical programs
do not need dependent types at runtime,
and with a bit of caution,
practical definitions can be kept agda2hs-supported.

\section{Discussion}
\label{sec:discussion}

We have presented an MVVM-inspired approach
for Agda/Haskell backends and C/C++ frontends;
as well as design patterns and primitives
that are contributions on their own,
including interruptible futures
and strategies for using Haskell built-ins in Agda.
We have also demonstrated that these contribute
to a use case that is viable compared to current solutions.

Note that many of our contributions
might have significance
beyond the scope of Agdalache.
Our future design can be reused
in any Haskell program,
regardless of whether it relies on Agda or even the FFI
(hence the Hackage library).
The Haskell dependency import approaches
are useful for agda2hs-based projects in general,
regardless of whether they use the FFI.
Finally, the basic idea of MVVM adaptation can be implemented
in other functional languages as well,
not just in Haskell or Agda.

The main target group of the framework
(i.e.\ developers writing Agda backends and C++ frontends
and using agda2hs)
is currently small, if not nonexistent.
Besides the above-described generality of most contributions,
we hope that the ideas presented here
will motivate people to try this path
of writing verified GUI programs.

\subsection*{Acknowledgements}

The author expresses his gratitude
to his supervisor Ambrus Kaposi
for assisting him throughout the years
while also giving him a free hand on the project.
Furthermore, he thanks his anonymous reviewers
for helping to position the article in the field
and make it of much better quality,
as well as Levente Lócsi and Gergő Szalay
for their useful suggestions and comments.
Finally, he sends his thoughts to his friends and family,
recalling the support they have provided
throughout the work.

This research has been supported by
the EKÖP-24 University Excellence Scholarship Program of
the Hungarian Ministry for Culture and Innovation
from the source of the
National Research, Development and Innovation Fund.

\bibliographystyle{ACM-Reference-Format}
\bibliography{bib.bib}

@article{gonthier,
    author = {Georges Gonthier},
    title = {Formal Proof — The Four-Color Theorem},
    journal = {Notices of the American Mathematical Society},
    year = {2008},
    url = {https://www.ams.org/notices/200811/tx081101382p.pdf},
    urldate = {2024-05-10},
}

@online{coq,
    title        = {The Coq Proof Assistant},
    author       = {{The~Coq~Team}},
    year         = {2024},
    urldate      = {2024-05-12},
    url          = {https://coq.inria.fr/},
}

@manual{agda-docs,
    author       = {Ulf Norell and others},
    title        = {The Agda User Manual, version 2.7.0.1},
    year         = {2024},
    urldate      = {2025-01-23},
    url          = {https://agda.readthedocs.io/en/v2.7.0.1/},
}

@manual{agda2hs-docs,
    author       = {Jesper Cockx and others},
    title        = {agda2hs Documentation},
    year         = {2022-2026},
    urldate      = {2026-03-21},
    url          = {https://agda.github.io/agda2hs/},
}

@online{norell-thesis,
    author       = {Ulf Norell},
    title        = {Towards a practical programming language based on dependent type theory},
    type         = {PhD thesis},
    institution  = {Chalmers University of Technology},
    year         = {2007},
    urldate      = {2024-05-12},
    url          = {http://www.cse.chalmers.se/~ulfn/papers/thesis.html},
}

@inproceedings{agda1-paper,
    author       = {Catarina Coquand and Thierry Coquand},
    title        = {Structured Type Theory},
    year         = {1999},
    booktitle    = {Proceedings of the
                       Workshop on Logical Frameworks and Meta-Languages},
    urldate      = {2024-05-12},
    url          = {https://www.eecs.uottawa.ca/~afelty/LFM99/CoquandCoquand.pdf},
}

@inproceedings{agda-emacs-paper,
    author       = {Catarina Coquand and Dan Synek and Makoto Takeyama},
    title        = {An Emacs-Interface for Type-Directed Support
                       for Constructing Proofs and Programs},
    year         = {2006},
    booktitle    = {Proceedings of the
                       European Joint Conferences on Theory and Practice of Software},
    volume      =  {2},
    urldate      = {2024-05-12},
    url          = {http://www.cse.chalmers.se/~coquand/emacs.pdf},
}

@inproceedings{agda2hs-paper,
author = {Cockx, Jesper and Melkonian, Orestis and Escot, Lucas and Chapman, James and Norell, Ulf},
title = {Reasonable Agda is correct Haskell: writing verified Haskell using agda2hs},
year = {2022},
isbn = {9781450394383},
publisher = {Association for Computing Machinery},
address = {New York, NY, USA},
url = {https://doi.org/10.1145/3546189.3549920},
doi = {10.1145/3546189.3549920},
booktitle = {Proceedings of the 15th ACM SIGPLAN International Haskell Symposium},
pages = {108–122},
numpages = {15},
location = {Ljubljana, Slovenia},
series = {Haskell 2022}
}

@unpublished{acorn-article,
  author = {Viktor Csimma},
  title = {Acorn – an agda2hs-compatible representation of exact real arithmetic},
  year = {2023},
  month = {11},
  location = {Budapest},
  urldate = {2024-04-30},
  url = {https://csimmaviktor.web.elte.hu/acorn.pdf},
}

@online{agda2hs,
    author       = {Jesper Cockx and others},
    title        = {agda2hs},
    subtitle     = {Compiling Agda code to readable Haskell},
    organization = {GitHub},
    year         = {{2020-2026}},
    urldate      = {2026-03-21},
    url          = {https://github.com/agda/agda2hs},
}

@online{agda-stdlib,
    author       = {Nils Anders Danielsson and others},
    title        = {The Agda standard library},
    organization = {GitHub},
    year         = {{2007-2024}},
    urldate      = {2024-05-10},
    url          = {https://github.com/agda/agda-stdlib},
}

@online{catch2,
    author       = {Martin Storsjö and others},
    title        = {Catch2},
    subtitle     = {A modern, C++-native, test framework for unit-tests, TDD and BDD},
    organization = {GitHub},
    year         = {{2010-2024}},
    urldate      = {2024-04-30},
    url          = {https://github.com/catchorg/Catch2},
}

@manual{ghc-docs-ffi,
    author = {{The~GHC~Team}},
    title        = {The Glasgow Haskell Compiler, 9.2.8.},
    subtitle     = {GHC User's Guide},
    chapter      = {6.17.},
    subtitle     = {Foreign Function Interface (FFI)},
    year         = {2020},
    urldate      = {2024-04-23},
    url          = {https://downloads.haskell.org/~ghc/9.2.8/docs/html/users_guide/exts/ffi.html},
}

@manual{base,
    author = {{The~GHC~Team}},
    title = {base-4.19.1.0},
    subtitle = {Basic libraries},
    organization = {Hackage},
    year = {2025},
    urldate = {2025-01-30},
    url = {https://hackage.haskell.org/package/base-4.19.1.0/docs/},
}

@manual{cabal-docs-what-cabal-does,
    author = {{The~Cabal~Team}},
    title        = {The Cabal User Guide, stable version},
    year         = {2024},
    subtitle     = {Cabal Explanation},
    chapter      = {1.},
    titleaddon   = {What Cabal does},
    urldate      = {2024-04-28},
    url          = {https://cabal.readthedocs.io/en/stable/cabal-context.html},
}

@online{stroustrup-faq,
    author       = {Bjarne Stroustrup},
    title        = {Bjarne Stroustrup's C++ Style and Technique FAQ},
    subtitle     = {Why doesn't C++ provide a "finally" construct?},
    year         = {2022},
    urldate      = {2024-05-10},
    url          = {https://www.stroustrup.com/bs_faq2.html#finally},
}

@online{malonzo-commit,
    author       = {Ulf Norell and others},
    title        = {Agda - commit d00cb80},
    organization = {GitHub},
    year         = {2008},
    urldate      = {2024-12-01},
    url          = {https://github.com/agda/agda/commit/d00cb80c6f924463318e1839a946df3a33e4aec1},
}

@manual{rocq-reference-manual,
    author       = {{The~Coq~Team}},
    title        = {The Coq Reference Manual, 8.20.0},
    organization = {Inria},
    year         = {2021},
    urldate      = {2024-12-01},
    url          = {https://coq.inria.fr/doc/V8.20.0/refman/index.html},
}

@manual{cpp23-draft,
    author       = {ISO/IEC},
    title        = {International Standard 14882:2024 – Programming Language C++ (working draft N4950)},
    year         = {2023},
    urldate      = {2025-01-23},
    url          = {https://open-std.org/JTC1/SC22/WG21/docs/papers/2023/n4950.pdf},
}

@online{agda2hs-compile-import,
    author       = {Heinrich Apfelmus and others},
    title        = {Better support for wrapping Haskell modules via postulate},
    subtitle     = {Issue \#316 at agda2hs},
    organization = {GitHub},
    year         = {2024},
    urldate      = {2024-12-01},
    url          = {https://github.com/agda/agda2hs/issues/316#issuecomment-2103042205},
}

@online{smalltt,
    author       = {András Kovács},
    title        = {AndrasKovacs/smalltt},
    organization = {GitHub},
    year         = {{2021}},
    urldate      = {2025-01-23},
    url          = {https://github.com/AndrasKovacs/smalltt/tree/3d4a20a6d80ac524325cf2d8d0a48095ded08eb6},
}

@unpublished{calc-article,
  author = {Viktor Csimma},
  title = {Agda for the masses: agda2hs-based libraries in real-world programs},
  year = {2024},
  month = {04},
  location = {Budapest},
  urldate = {2024-04-30},
  url = {https://csimmaviktor.web.elte.hu/calc.pdf},
}

@article{veriffi,
author = {Korkut, Joomy and Stark, Kathrin and Appel, Andrew W.},
title = {A Verified Foreign Function Interface between Coq and C},
year = {2025},
issue_date = {January 2025},
publisher = {Association for Computing Machinery},
address = {New York, NY, USA},
volume = {9},
number = {POPL},
url = {https://doi.org/10.1145/3704860},
doi = {10.1145/3704860},
journal = {Proc. ACM Program. Lang.},
month = jan,
articleno = {24},
numpages = {31}
}

@book{cpp-evolution,
    author = {Bjarne Stroustrup},
    title = {The Design and Evolution of C++},
    publisher = {Addison-Wesley},
    address = {New York},
    isbn = {ISBN 978-0-201-54330-8},
    year = {1994},
}

@manual{old-future,
    author = {Edward Kuklewicz},
    title = {future},
    subtitle = {Supposed to mimics and enhance proposed C++ "future" features},
    organization = {Hackage},
    year = {{2009}},
    urldate = {2025-09-07},
    url = {https://hackage.haskell.org/package/future},
}

@manual{cfuture,
    author = {Viktor Csimma},
    title = {cfuture},
    subtitle = {A Future type that is easy to represent and handle in C/C++},
    organization = {Hackage},
    year = {{2025}},
    urldate = {2025-09-07},
    url = {https://hackage.haskell.org/package/cfuture},
}

@inproceedings{future-1977,
author = {Baker, Henry C. and Hewitt, Carl},
title = {The incremental garbage collection of processes},
year = {1977},
isbn = {9781450378741},
publisher = {Association for Computing Machinery},
address = {New York, NY, USA},
url = {https://doi.org/10.1145/800228.806932},
doi = {10.1145/800228.806932},
booktitle = {Proceedings of the 1977 Symposium on Artificial Intelligence and Programming Languages},
pages = {55–59},
numpages = {5}
}

@online{rewrite-rules-pr,
    author       = {Viktor Csimma},
    title        = {Option for user-defined rewrite rules given in a config file},
    subtitle     = {agda2hs - pull request 189},
    organization = {GitHub},
    year         = {2023},
    urldate      = {2025-09-11},
    url          = {https://github.com/agda/agda2hs/pull/189},
}

@online{mvvm-2005,
    author       = {John Gossman},
    title        = {Introduction to Model/View/ViewModel pattern for building WPF apps},
    year         = {2005},
    month        = {October},
    organization = {Microsoft},
    url          = {https://learn.microsoft.com/en-us/archive/blogs/johngossman/introduction-to-modelviewviewmodel-pattern-for-building-wpf-apps},
    urldate      = {2026-03-08}
}

@Inbook{mvvm-book,
author="Garc{\'i}a, Ra{\'u}l Ferrer",
title="MVVM: Model--View--ViewModel",
bookTitle="iOS Architecture Patterns: MVC, MVP, MVVM, VIPER, and VIP in Swift",
year="2023",
publisher="Apress",
address="Berkeley, CA",
pages="145--224",
isbn="978-1-4842-9069-9",
doi="10.1007/978-1-4842-9069-9_4",
url="https://doi.org/10.1007/978-1-4842-9069-9_4"
}

@manual{android-threads,
    author       = {Google},
    title        = {Android Developers},
    subtitle     = {Processes and threads overview},
    year         = {2024},
    urldate      = {2025-09-11},
    url          = {https://developer.android.com/guide/components/processes-and-threads#Threads}
}

@article{compcert-paper,
author = {Xavier Leroy},
title = {Formal verification of a realistic compiler},
year = {2009},
issue_date = {July 2009},
publisher = {Association for Computing Machinery},
address = {New York, NY, USA},
volume = {52},
number = {7},
issn = {0001-0782},
url = {https://doi.org/10.1145/1538788.1538814},
doi = {10.1145/1538788.1538814},
journal = {Commun. ACM},
month = jul,
pages = {107–115},
numpages = {9}
}

\label{lastpage01}

\end{document}